\documentclass{elsart}
\usepackage{graphicx,amssymb}
\journal{Chemical Physics Letters}
\begin{document}
\begin{frontmatter}
\title{Possible EIT-like effects in strong-field photodissociation of carbon disulphide}
\author[tifr]{Firoz A. Rajgara},
\author[tifr]{Deepak Mathur\corauthref{cor}},
\corauth[cor]{Corresponding author.}
\ead{atmol1@tifr.res.in}
and
\author[rri]{Hema Ramachandran}
\address[tifr]{Tata Institute of Fundamental Research, 1 Homi Bhabha Road,
Mumbai 400 005, India}
\address[rri]{Raman Research Institute, Sadashiv Nagar, Bangalore 560 080, India}

\begin{abstract}
CS$_2$ molecules are spatially aligned upon irradiation by intense (1-100 TW cm$^{-2}$), 35 ps pulses of 512 nm or 355 nm light. When both colours are simultaneously present, spatial alignment disappears. We draw analogy with weak-field illumination of an atom by two colours wherein EIT (electromagnetically induced transparency) inhibits absorption by virtue of dipoles not being induced in two-colour fields. In the strong field scenario, molecular alignment is a consequence of a dipole being induced in the direction of the applied electric field. It follows, therefore, that when dipoles are not induced, no spatial alignment is to be expected.    
\end{abstract}
\end{frontmatter}

\section{Introduction}

Electromagnetically induced transparency (EIT) is a well established quantum-optical phenomenon wherein a normally absorbing optical medium is rendered transparent upon its irradiation by a two-colour field \cite{shen,knight,agarwal,marangos}. The much-studied V system of EIT involves three atomic levels, as indicated in Fig 1. A transition from the ground state $\Psi$ to an excited state $\Psi$1 may be made by the absorption of a photon at $\omega_1$, or a transition to $\Psi$2 by absorption of an $\omega_2$ photon. The  simultaneous presence of fields at both frequencies $\omega_1$ and $\omega_2$ causes a quantum interference of the two absorption processes, and for a certain narrow range of frequencies and strength of lasers, complete cancellation of transition amplitudes renders the medium transparent to the light.  This phenomenon, which has attracted considerable recent interest, has invariably been studied under application of relatively weak fields, typically few mW cm$^{-2}$. We report here what we believe might be the first observation of an EIT-like phenomenon in the strong-field (TW cm$^{-2}$), two-colour, angle-resolved photo-dissociation of a molecule, CS$_2$. 

Strong non-resonant laser fields are known to cause spatial alignment of linear molecules along the optical field polarization direction \cite{herschbach,mathur}.  However, we show here a suppression of this effect by the simultaneous application of two strong fields at two different colours, each of which, when applied individually, does indeed produce spatial alignment. 

Before describing our results, it is of interest to note that in normal (weak field) illumination of an atom by 2 colours, upon the occurrence of EIT, a particular absorption is inhibited. In other words, dipoles cease to be induced when the two colours are simultaneously present. In the strong field scenario, spatial alignment of a molecule is a consequence of a dipole being induced in the direction of the applied electric field. It follows, therefore, that when dipoles are not induced, no spatial alignment is to be expected.  If that be the case, might not an EIT-like phenomenon then inhibit spatial alignment of molecules in the strong field regime?  We note that light intensities that are typically used in strong field experiments are usually of more than sufficient magnitude to also give rise to multiple electron ejection from the aligned molecule. The experimental signature of spatial alignment is, therefore, readily available in the form of the anisotropy in the angular distribution of fragment ions that result from Coulomb explosion of a multiply charged molecule. We therefore investigated this feature in our experiments.

\section{Experimental Method} 
 
Our experimental observations of possible EIT-effects in the strong-field regime involved angle-resolved measurement of the products of dissociative ionization of CS$_2$ upon irradiation by 35 ps-long pulses of light at 532 nm and 355 nm, of intensity values in the 1-100 TW cm$^{-2}$ range. The main features of the apparatus that we used that are pertinent to the present set of experiments are summarized in Fig. 2. Optical field induced ionization of a thermal beam of initially randomly oriented CS$_2$ molecules was achieved in an ultrahigh vacuum chamber (pressure $<$3$\times$10$^{-9}$ Torr). CS$_2^+$ and fragment ions were made to pass through a quadrupole mass spectrometer (QMS) preceded by an electrostatic energy analyzer. Energy- and mass-selected ions were detected by a single particle multiplier. By varying the polarization direction of the incident laser radiation, angle-resolved measurement of the ion products could be made. The use of a QMS provided us with the distinct advantage of being able to measure pristine angular distributions, without recourse to use of electrostatic extraction fields that often lead to distortion of such distributions. 

The scheme that we adopted for inducing ionization using two different colours is indicated in Fig. 2.  The fundamental of the Nd:YAG laser (1064 nm) and its second harmonic (532 nm) were split into two beams, one of which was used to generate a relatively weak ($<$0.6 mJ) third harmonic at 355 nm using a DKDP crystal. A strong (12 mJ) beam at 532 nm was made to pass through an optical delay that was introduced using a retro-reflector mounted on a translation stage. Temporal overlap in these two beams was achieved using a mirror placed before the laser-molecule interaction zone. Initially, the beam used to generate 355 nm light was slightly misaligned vis-\'a-vis the DKDP crystal and only the 532 nm component was used. The two 532 nm beams were spatially overlapped in non-collinear geometry (angle $<$4 ) onto a thin film of R6G dye doped in a gel. In the case of R6G, there is significant absorption at the exciting wavelength and this leads to a population grating. The first-order diffracted signal that arises due to the third-order nonlinearity depends on the temporal overlap of the two beams \cite{eichler,dharmadhikari}. The diffracted signal was detected by a photodiode (PD1); temporal fluctuations in the laser energy were monitored using a second photodiode (PD2). The photodiode signals were acquired using a digital oscilloscope (DSO). The maximum value of this signal as a function of temporal delay provided an excellent indicator of maximum overlap between the two beams. Having ensured maximum overlap, the mirror M1 was removed and the DKDP crystal was properly realigned such that both 355 nm and 532 nm beams were nearly collinear and focused onto the CS$_2$ gas jet. 

The two-colour light was focused using a biconvex lens, of focal length 20 cm, placed in an ultrahigh vacuum chamber capable of being pumped down to a base pressure of 3$\times$10$^{-11}$ Torr. In our measurements, typical laser intensities at the focal spot were in the range 1-100 TW cm$^{-2}$; typical operating pressures were in the 10$^{-9}$ Torr range, low enough for us to be able to safely ignore space charge effects on our spectra. The spatial distribution of ions produced in the focus was examined by inserting apertures of different sizes before the detector so as to spatially limit the interaction volume being sampled. The size of the aperture to be used depends on the Rayleigh range. By scanning a knife edge along the cross section of the laser beam using a computer-driven translation stage we deduced the radius of the focused spot to be 24 $\mu$m; the Rayleigh range was $\sim$2.4 mm. In the present series of measurements we used circular apertures, of 2 mm diameter, centered about the focal point so that only the Rayleigh range was sampled. A combination of halfwave plate and polarizer was used to obtain the desired polarization at constant intensity.  The polarization direction was rotated in steps of 1-4$^o$. The shot-to-shot reproducibility of the laser was monitored on-line by a fast photodiode coupled to the DSO and in the course of most of the measurements, the shot-to-shot laser intensity variation was $\pm$5\%.

\section{Results and Discussion}

We depict in Fig. 3 typical angular distributions of S$^+$ fragments that result from optical-field dissociative ionization of CS$_2$ using 532 nm and 355 nm light, each applied singly. As is clearly seen, the ion signal falls substantially as the angle deviates toward 90$^{o}$. This has been interpreted and modeled \cite{pendular,alignment} as being indicative of spatial alignment of the linear CS$_2$ molecule along the {\bf E}-vector of the applied light. A dramatically different S$^+$ angular distribution is obtained when 532 nm and 355 nm light are simultaneously applied. This is depicted in Fig. 4: the angular asymmetry is denoted by the ratio of the S$^+$ signal measured with laser polarization in the 0$^{o}$ direction (parallel to the QMS axis) to that measured when the polarization direction is orthogonal. In these measurements, the intensity of the 355 nm light was kept at low enough values to ensure that optically induced ionization of CS$_2$ could not be caused by the 355 nm light alone (threshold $\sim$10$^{12}$ W cm$^{-2}$); data depicted in Fig. 4 pertains to the intensity of the 532 nm light being kept fixed at 8$\times$10$^{13}$ W cm$^{-2}$ intensity. The anisotropy parameter (S$^+_{parallel}$ / S$^+_{perpendicular}$) decreases with intensity of the 355 nm light, implying that, at the highest 355 nm intensity, there ceases  to be distinction between the 0$^{o}$ and 90$^{o}$ directions, that is, the spatial alignment of molecules reduces when sufficiently intense 355 nm light is also applied. This interesting and unexpected observation leads us to make an analogy with the much-studied V-system of EIT (Fig. 1). 

Strong-field, angle-dependent photo-dissociation of the type that is manifested in Fig. 3 is, in the picosecond regime, a two-step process, the first being the alignment of the molecule and the second the fragmentation. In our experiment, it is the alignment that appears to be inhibited; the fragmentation occurs whether one laser field is present, or several. We therefore concentrate on the process of alignment. The process of ``off-resonantly inducing a dipole" is actually a strong-field effect, where the sensitivity to frequency is essentially absent. The presence of a high-intensity laser field at 532 nm causes a distortion of the intramolecular electronic charge cloud \cite{kvl1}.  The electronic wavefunction acquires the symmetry of the applied field, that is, the probability density maximizes along the direction of the polarization of the applied electric field \cite{kvl1} and the electrons oscillate at the frequency of the applied field. This is the induced dipole and its alignment to the field. In similar fashion, light at 355 nm would cause a distortion of the wavefunction to $\Psi$2, which too has the symmetry of the applied field. As the 355 nm light is weak, this can effectively induce dipoles only resonantly. In the simultaneous presence of both fields, however, the probability amplitudes for the two processes appear to interfere, causing cancellation. This is a possible reason as to why when both lasers are simultaneously present, and even though they are polarized along the same direction, re-orientation of the CS$_2$ molecule is absent. Unlike low-field EIT, this cancellation is not expected to be spectrally narrow in the strong field case as the intense light, even if non-resonant, induces dipoles to a significant extent. However, like in conventional EIT, the process is likely to depend on the relative strength of the two lasers, in the same manner that the contrast of interference fringes depends on the strength of the two sources. The intensity dependence of the ``contrast" in our case is amply evident in Fig. 4.
It should be noted here that while the two fields are pulsed, the pulses are temporally matched as they are derived from the same source - the fundamental of our Nd:YAG laser. Their envelopes are therefore similar, and the pulses at the two colours overlap spatially and temporally at the sample. The duration of the pulses is shorter than the decoherence time.  We estimate that our pulses are phase matched to within 30 ps. The role of phase in strong-field molecular dynamics has been highlighted in several reports that address issues of control of the electronic wavepacket \cite{leone}. In our study, the role of phase in determining the motion of nuclei remains to be elucidated and deserves further work. However, it appears to us that the suppression of the induced dipolar alignment is unlikely to be directly attributable to phase changes. In our experiments, the suppression of alignment is directly related to the magnitude of the optical field; as our light intensity is altered no phase change is expected to occur. 

How do we ascribe an EIT-like process specifically to the CS$_2$ molecule?  An experimental study of dipole-allowed and dipole-forbidden electronic excitations in CS$_2$ \cite{mk} reveals a plethora of possible electronic states that might participate in an EIT-like process. Of these, the lowest lying state involves excitation from the linear ground electronic state $X~^1\Sigma_g$ to the bent state $a^3$A$_2$, whose vibrational levels span the energy region between 2.5 - 3.9 eV above the zeroth vibrational level of the ground state \cite{mk}. This state is accessible by a single 355 nm photon. The singlet-triplet transition is allowed in this case because the spatial orientation of the molecule provides the conservation of angular momentum. Singlet-singlet transitions can also occur via two-photon absorptions to a number of linear excited states that lie in the energy region beyond 7 eV. Our intensity dependent experiments reveal that it is the former (one-photon) excitation that occurs with high propensity (see inset to Fig. 3b). In the context of Fig.1, $\Psi$2 therefore corresponds to the $a^3$A$_2$ state of CS$_2$. We remark here that ionization of the molecule can occur by multiphoton absorption by the states $\Psi$1 and/or $\Psi$2. 

The ground electronic state of the CS$_2$ molecule has the electronic configuration
\begin{eqnarray*}
(core)^{22}(5{\sigma_g})^2(4{\sigma_u})^2(6{\sigma_g})^2
(5{\sigma_u})^2(2{\pi_u})^4(2{\pi_g})^4
\end{eqnarray*}
Ejection of one electron from the outermost $\pi_g$ orbital leads to the formation of CS$_2^+$ in the ground  electronic state,$X^2\Pi_g$, while removal of an electron from 2$\pi_g$, 5$\sigma_u$ and 6$\sigma_g$ 
orbitals gives rise to excited $A$, $B$, and $C$ electronic states, respectively, of CS$_2^+$. Fragment ions like S$^+$ are not produced upon direct ionization of neutral CS$_2$ since the appropriate Franck-Condon factors indicate that the dissociation continua of the $X$, $A$ and $B$ electronic states are not accessible by vertical ionization from the CS$_2$ ground state. This was established a long time ago by the absence of any peak in CS$_2$ photoelectron spectra \cite{Brundle} near the 14.81 eV appearance threshold for S$^+$, and also by photoion-photoelectron coincidence measurements \cite{Brehm}.  The ionic state, $C$, lies 16.2 eV above the CS$_2$ ground state, above the dissociation limit for S$^+$ + CS (and S + CS$^+$) and, hence, fully predissociates. A higher-energy band, designated the $D$ state in photoionization experiments \cite{Brundle}, has a broad featureless peak centered around 17 eV, spreading over 1.2 eV; this band is formed as a result of the ejection of an electron from the 4$\sigma_u$ orbital of CS$_2$ and is one of the multielectron states (MES) that are so important in any quantumchemical description of the CS$_2$ molecule \cite{Mathur}. The CS$^+$ ($^2\Sigma^+$) + S ($^3P_g$) and S$^+$ ($^4S_u$) + CS ($^1\Sigma^+$) dissociation limits are 0.42 and 1.4 eV below the $C$ state. The next S$^+$ + CS dissociation channel lies at 16.65 eV, 0.45 eV above the $C$ state. So the $C$ and MES states can, on purely energetic grounds, lead to the formation of both the CS$_2^+$ molecular ion as well as to fragment ions like S$^+$. 

We also carried out experiments on the valence-isoelectronic molecule CO$_2$, which affords no option for a singlet photon excitation process with 355 nm light \cite{mk}: no EIT-like process was observed in CO$_2$, strongly indicating that the presence of the low-lying triplet excited state in CS$_2$ is a necessary ingredient for the EIT-like process if one of the light beams is weak, (like the 355 nm light in our case).

\section{Concluding Remarks}

We have irradiated CS$_2$ molecules with intense (1-100 TW cm$^{-2}$), 35 ps-long pulses of 512 nm or 355 nm light that is linearly polarized. Spatial alignment of the linear S-C-S molecule is observed in both cases, as exemplified by the anisotropic angular distributions of S$^+$ fragment ions that are produced upon optical-field-induced dissociative ionization. However, when both colours are simultaneously present, spatial alignment disappears. This situation is akin to that observed upon weak-field illumination of an atom by two colours wherein an electromagnetically induced transparency can inhibit absorption by virtue of dipoles not being induced in two-colour fields. Strong field molecular alignment is also a consequence of a dipole being induced in the direction of the polarization vector. It follows, therefore, that when dipoles are not induced, lack of spatial alignment may not be unexpected. 

Providing an exact theoretical treatment of this possible EIT-like phenomenon is a formidable task. Usual treatments of EIT consider much lower applied fields, typically a few mW cm$^{-2}$. The steady state density matrix for  the three-level atom in the interaction picture is obtained by solving the  Liouville equation.   However, for the case under discussion, the usual approaches are not readily applicable - the strong field drastically modifies the level structure and also causes redistribution of population among states. Most assumptions made in traditional treatments are not valid at intense fields; this calls for new theoretical approaches. One such effort is that of  Tan {\it et al.} \cite{tan}, involving the concept of self-regulated partitioning of the interaction. The interaction Hamiltonian is partitioned into two, one part being responsible for modifying the states to a new set of basis states and the other for excitation of the modified states. However, the  exact proportion of partitioning depends on the process under study and presents obvious difficulties in the strong field regime. It is hoped that our experimental work will stimulate theoretical efforts in this direction. 

\section{Acknowledgment} 
We gratefully acknowledge A.K. Dharmadhikari's skillful help in achieving and quantifying the spatial and temporal overlap of the two-colour beams used in these experiments.

\newpage
\begin{figure}
\caption{\label{f1}(a) Schematic energy levels showing a V-system formed by the ground state $\Psi$ and two excited states $\Psi$1 and $\Psi$2. Also shown is a continuum of levels (hatched area). (b) For a level structure similar to that shown in (a), the typical absorption of continuous wave laser light of frequency $\omega$, scanned about the value $\omega_2$, is shown in the upper panel, when that laser alone is present. The lower panel shows the appearance of a transparency window at $\omega_2$ in the absorption of the same laser, under the EIT condition which may be satisfied with the simultaneous presence of a second laser at frequency $\omega_1$.}
\end{figure}

\begin{figure}
\caption{\label{f2}Schematic representation of the apparatus used in the present experiments (see text). M: mirrors. The angle between the 2$\omega$ and 3$\omega$ beams as they enter the vacuum chamber is exaggerated; in practice it is very small.} 
\end{figure}

\begin{figure}
\caption{\label{f3}Angle-resolved intensity of S$^+$ fragment ions that result from dissociative ionization of CS$_2$ upon irradiation by a) 532 nm light and b) 355 nm light. In both cases the peak intensity was 10$^{14}$ W cm$^{-2}$. The inset to b) shows the linear variation of the asymmetry parameter as a function of 355 nm light intensity (see text). 6 mJ of 355 nm light corresponded to 10$^{14}$ W cm$^{-2}$ intensity.} 
\end{figure}

\begin{figure}
\caption{\label{f4}Asymmetry parameter (ratio of S$^+$ signal obtained with parallel polarization to that obtained with perpendicular polarization) measured in a two-colour field as a function of the incident 355 nm laser energy. An asymmetry parameter of zero indicates isotropic S$^+$ angular distribution. The 532 nm laser energy was kept fixed at 11 mJ in these measurements. 0.6 mJ of 532 nm light corresponded to 0.6$\times$10$^{13}$ W cm$^{-2}$.}
\end{figure}

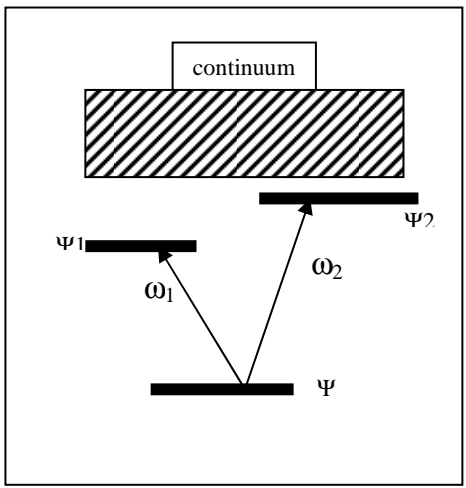
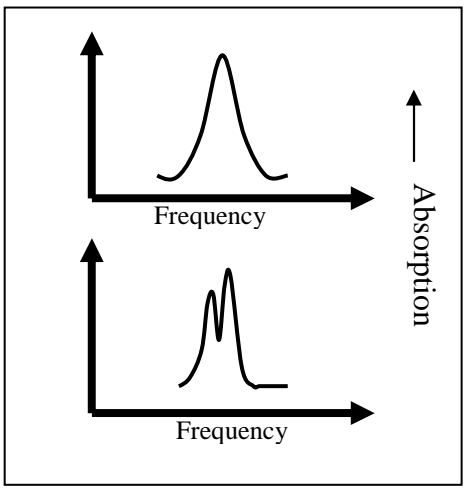
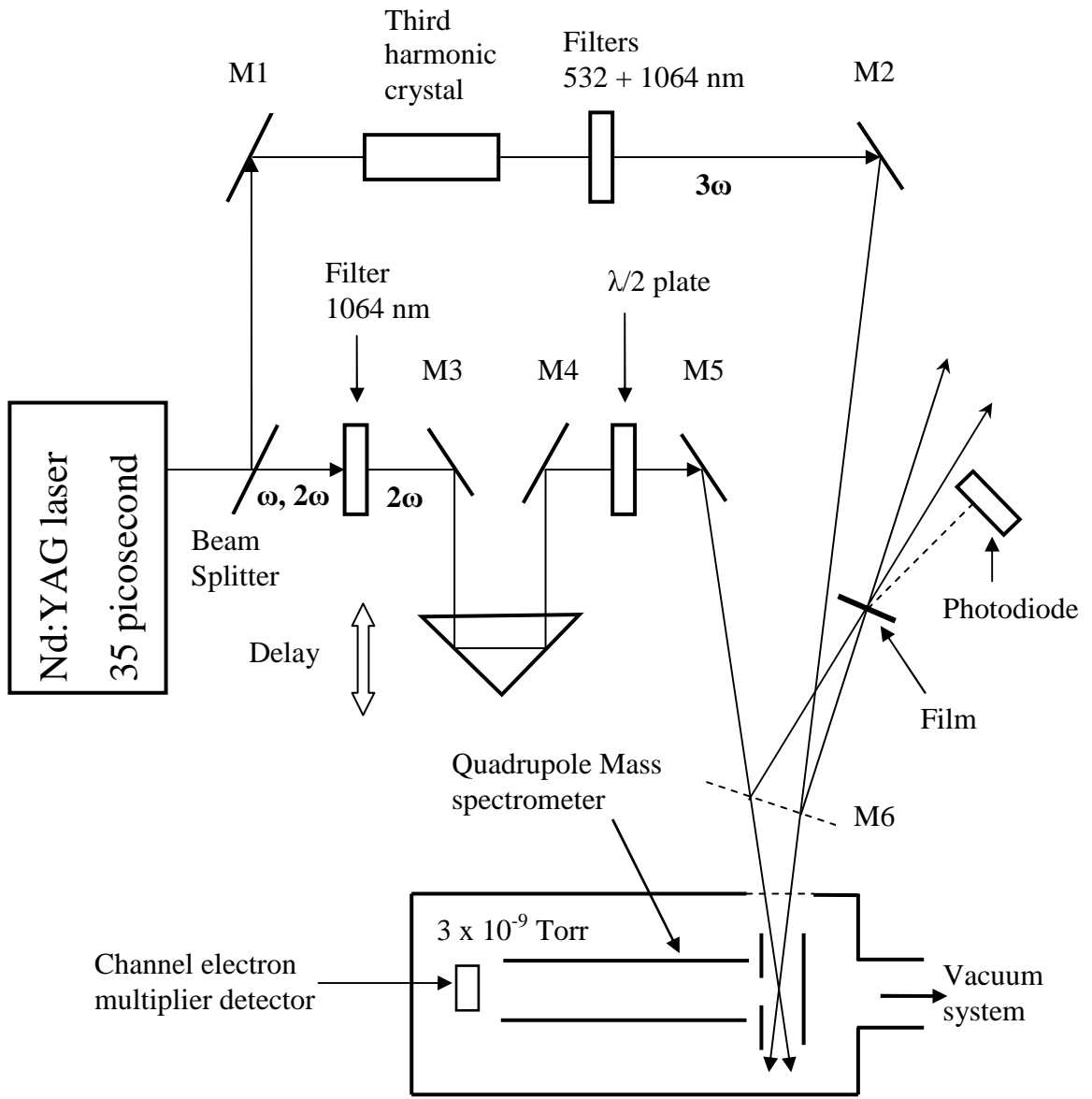
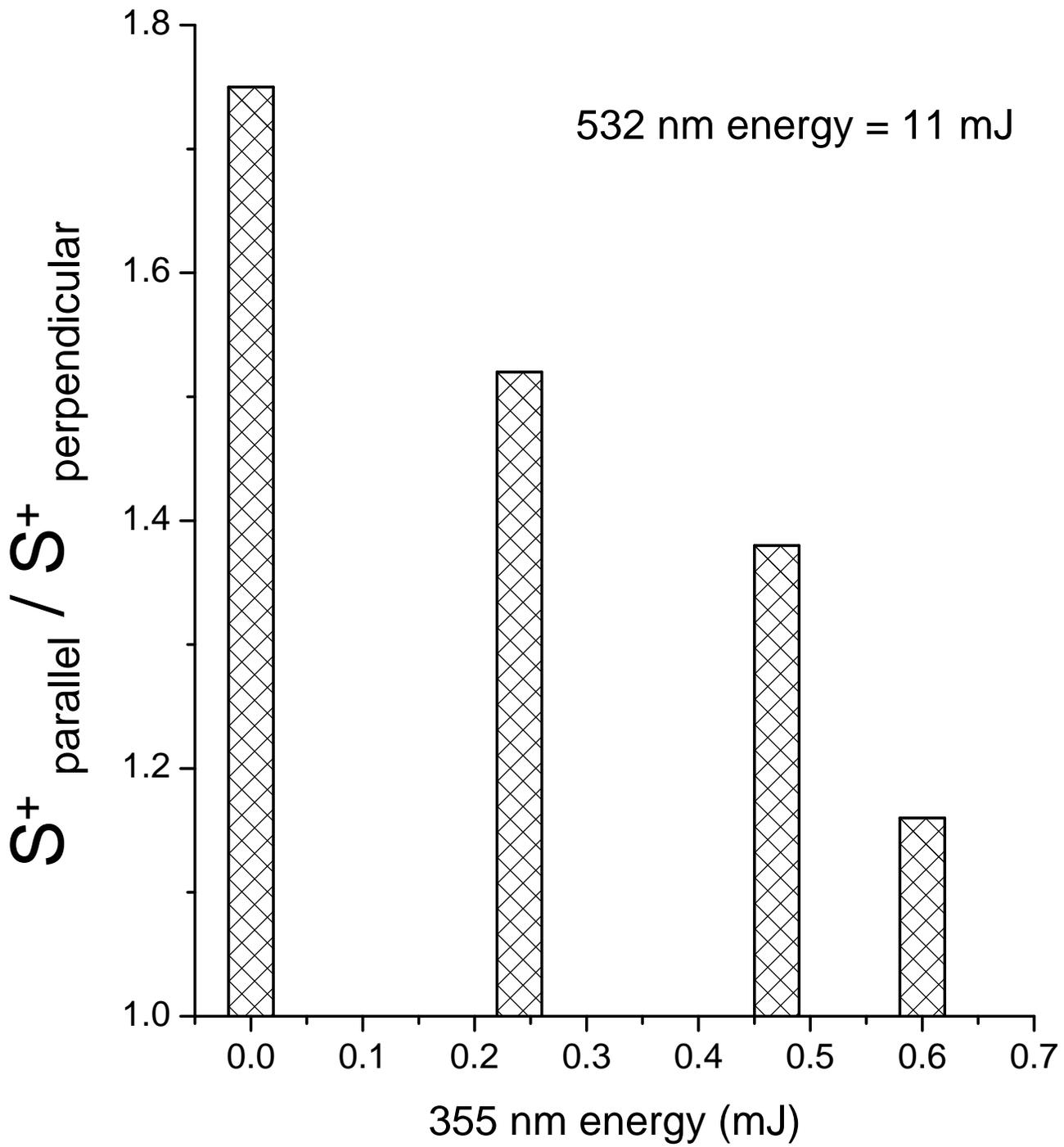
\end{document}